\documentclass[slac_one]{revtex4}
\usepackage{graphicx}
\usepackage{fancyhdr}
\newcommand{\Kspipi}{\ensuremath{B^{0} \rightarrow K^{0}_{S} \pi^{+} \pi^{-}}}
\newcommand{\Kspiz}{\ensuremath{B^{0} \rightarrow K^{0}_{S} \pi^{0}}}
\newcommand{\Klpiz}{\ensuremath{B^{0} \rightarrow K^{0}_{L} \pi^{0}}}
\newcommand{\Kzpiz}{\ensuremath{B^{0} \rightarrow K^{0} \pi^{0}}}
\newcommand{\Kskk}{\ensuremath{B^{0} \rightarrow K^{0}_{S} K^{+} K^{-}}}

\newcommand{\epem}{\ensuremath{e^{+} e^{-}}}
\newcommand{\Ups}{\ensuremath{\Upsilon (4S)}}
\newcommand{\BBbar}{\ensuremath{B \bar B}}

\newcommand{\BzBzb}{\ensuremath{B^{0} \bar B^{0}}}
\newcommand{\qqbar}{\ensuremath{q \bar q}}

\newcommand{\Mbc}{\ensuremath{M_{\rm bc}}}
\newcommand{\De}{\ensuremath{\Delta E}}

\newcommand{\LR}{\ensuremath{{\cal R}_{\rm S/B}}}

\newcommand{\pip}{\ensuremath{\pi^{+}}}
\newcommand{\pim}{\ensuremath{\pi^{-}}}

\newcommand{\Kp}{\ensuremath{K^{+}}}
\newcommand{\Km}{\ensuremath{K^{-}}}
\newcommand{\Ks}{\ensuremath{K^{0}_{S}}}
\newcommand{\Kl}{\ensuremath{K^{0}_{L}}}

\newcommand{\chic}{\ensuremath{\chi_{c0}}}

\newcommand{\Kspm}{\ensuremath{K^{*\pm}(892)}}
\newcommand{\Kstarp}{\ensuremath{K^{*+}_{0}(1430)}}

\newcommand{\Kstarpm}{\ensuremath{K^{*\pm}_{0}(1430)}}
\newcommand{\rhoz}{\ensuremath{\rho^{0}(770)}}
\newcommand{\fz}{\ensuremath{f_{0}(980)}}
\newcommand{\ftwo}{\ensuremath{f_{2}(1270)}}
\newcommand{\fX}{\ensuremath{f_{X}(1300)}}
\newcommand{\Bz}{\ensuremath{B^{0}}}
\newcommand{\Bzb}{\ensuremath{\bar B^{0}}}

\newcommand{\Brec}{\ensuremath{B^{0}_{\rm Rec}}}
\newcommand{\Btag}{\ensuremath{B^{0}_{\rm Tag}}}

\newcommand{\Dt}{\ensuremath{\Delta t}}
\newcommand{\taub}{\ensuremath{\tau_{\Bz}}}
\newcommand{\Dmd}{\ensuremath{\Delta m_{d}}}

\newcommand{\Acp}{\ensuremath{{\cal A}_{CP}}}
\newcommand{\Scp}{\ensuremath{{\cal S}_{CP}}}

\newcommand{\phione}{\ensuremath{\phi^{\rm eff}_{1}}}

\begin{document}

\title{Measurements of CKM angle $\phi_{1}$ with charmless penguins at Belle} 

%

\author{J.~Dalseno for the Belle Collaboration}
\affiliation{High Energy Accelerator Research Organization (KEK), Tsukuba, JAPAN}

\begin{abstract}
  We present measurements of time-dependent $CP$ violation parameters in \Kzpiz, \Kspipi\ and \Kskk\ decays. The latter two are extracted using time-dependent Dalitz plot analyses. These results are obtained from a large data sample that contains $657 \times 10^{6}$ \BBbar\ pairs collected at the \Ups\ resonance with the Belle detector at the KEKB asymmetric-energy \epem\ collider.
\end{abstract}

\maketitle

\thispagestyle{fancy}

\section{INTRODUCTION} 
$CP$ violation in the Standard Model (SM) arises from an irreducible complex phase in the Cabibbo-Kobayashi-Maskawa (CKM) quark-mixing matrix~\cite{C,KM}. Of recent interest is $CP$ violation in $b \rightarrow q \bar q s$ transitions which proceeds by loop diagrams that may be affected by new particles in various extensions of the SM. Furthermore, the $CP$ asymmetries in $b \rightarrow q \bar q s$ transitions are predicted in the SM to be slightly higher than those observed in $b \rightarrow c \bar c s$ transitions. However, current experimental measurements~\cite{HFAG} tend to be lower than those for $b \rightarrow c \bar c s$ transitions motivating more precise experimental determinations.

The decay of the \Ups\ produces a \BBbar\ pair of which one (\Brec) may be fully reconstructed while the other (\Btag) may reveal its flavour. The proper time interval between \Brec\ and \Btag\ is defined as $\Dt \equiv t_{\rm Rec} - t_{\rm Tag}$ and from coherent \BBbar\ production in the \Ups\ decay, the time-dependent decay rate for a quasi-two-body mode when \Btag\ possesses flavour $q$ (\Bz: $q=+1$, \Bzb: $q=-1$), is given by~\cite{Sanda}
\begin{equation}\label{eq_tcpv}
  |A(\Dt, q)|^{2} = \frac{e^{-|\Dt|/\taub}}{4\taub} \biggl[ 1 + q(\Acp \cos \Dmd \Dt + \Scp \sin \Dmd \Dt) \biggr],
\end{equation}
where \taub\ is the \Bz\ lifetime and \Dmd\  is the \BzBzb\ mass difference. This assumes no $CP$ violation in mixing, $|q/p|=1$, and that the  \BzBzb\ lifetime difference is negligible. The parameter, \Acp, denotes the direct $CP$ violating component and \Scp\ represents mixing-induced $CP$ violation. For time-dependent Dalitz plot analyses, the time-dependent decay rate is written as
\begin{equation}
  |A(\Dt, q)|^{2} = \frac{e^{-|\Dt|/\taub}}{4\taub} \biggl[ (|A|^{2} + |\bar A|^{2}) - q(|A|^{2} - |\bar A|^{2}) \cos \Dmd \Dt + 2q \Im (\bar A A^{*}) \sin \Dmd \Dt \biggr].
\end{equation}
The Dalitz-dependent amplitudes, $A$, can be written in the isobar approximation as a superposition of intermediate decay channels, $i$,
\begin{equation}\label{eq_sigmod}
  A(s_{+}, s_{-}) =\sum_{i} a_{i}F_{i}(s_{+}, s_{-}), \;\;\;\; \bar A(s_{-}, s_{+}) =\sum_{i} \bar a_{i} \bar F_{i}(s_{-}, s_{+}),
\end{equation}
where $a_{i} \equiv a_{i}(1 + c_{i})e^{i(b_{i} + d_{i})}$ for $A$ and $\bar a_{i} \equiv a_{i}(1 - c_{i})e^{i(b_{i} - d_{i})}$ for $\bar A$ are complex coefficients describing the relative magnitudes and phases between the decay channels. The form factors, $F_{i}(s_{+}, s_{-})$, depend on the Dalitz plot coordinates, $s_{\pm}$, and describe the invariant mass and angular distribution probabilities. For a $CP$ eigenstate, $i$, the time-dependent $CP$ violation parameters can be calculated as
\begin{equation}
  \Acp(i) \equiv \frac{|\bar a_{i}|^{2}-|a_{i}|^{2}}{|\bar a_{i}|^{2}+|a_{i}|^{2}} =  \frac{-2c_{i}}{1+c_{i}^{2}}, \;\; {\rm and} -\eta_{i}\Scp(i) \equiv \frac{-2\Im(\bar a_{i}a^{*}_{i})}{|a_{i}|^{2} + |\bar a_{i}|^{2}} = \frac{1-c^{2}_{i}}{1+c^{2}_{i}}\sin 2 \phione(i)
\end{equation}
and $\phione(i) \equiv {\arg(a_{i} \bar a^{*}_{i})}/{2} = d_{i}$ is directly accessible as a fit parameter. Consequently, both $\Acp(i)$ and $\Scp(i)$ are restricted to reside in the physical region. The relative fraction of each component can be calculated with,
\begin{equation}
  f_{i} = \frac{(|a_{i}|^{2} + |\bar a_{i}|^{2}) \int F_{i}(s_{+}, s_{-}) F^{*}_{i}(s_{+}, s_{-}) ds_{+} ds_{-}}{\int  (|{\cal A}|^{2} + |\bar {\cal A}|^{2}) ds_{+} ds_{-}}.
\end{equation}

\section{DATASET, DETECTOR AND BASIC ANALYSIS TECHNIQUE}
These measurements of $CP$ violating parameters are based on $657 \times 10^6 \; B \bar B$ pairs collected  with the Belle detector~\cite{Belle} at the KEKB asymmetric-energy $e^+e^-$ ($3.5$ on $8~{\rm GeV}$) collider~\cite{KEKB}. Operating with a peak luminosity that exceeds $1.7\times 10^{34}~{\rm cm}^{-2}{\rm s}^{-1}$, the collider produces the \Ups\ resonance ($\sqrt{s}=10.58$~GeV) with a Lorentz boost of $\beta\gamma=0.425$, opposite to the positron beam direction, $z$, which usually decays into a \BBbar\ pair.

Reconstructed $B$ candidates are described with two kinematic variables: the beam-constrained mass, $\Mbc \equiv \sqrt{(E^{\rm CMS}_{\rm beam})^{2} - (p^{\rm CMS}_{B})^{2}}$ and the energy difference, $\De \equiv E^{\rm CMS}_{B} - E^{\rm CMS}_{\rm beam}$ where $E^{\rm CMS}_{\rm beam}$ is the beam energy and $E^{\rm CMS}_{B}$ ($p^{\rm CMS}_{B}$) is the energy (momentum) of the $B$ meson all evaluated in the centre-of-mass system (CMS). The dominant background in the reconstruction of \Brec\ is from continuum ($\epem \rightarrow \qqbar$) events. Since their topology tends to be jet-like in contrast to the spherical \BBbar\ decay, continuum can be suppressed with a Fisher discriminant based on modified Fox-Wolfram moments~\cite{SFW}. This discriminant is combined with the polar angle of the $B$ candidate in the CMS to form a likelihood ratio, \LR, which separates continuum from \BBbar\ events.

Since the $B^0$ and $\bar{B}^0$ mesons are approximately at rest in the \Ups\ Enter-of-Mass System (CMS), the difference in decay time between the $B \bar B$ pair, $\Delta t$, can be determined from the displacement in $z$ between the final state decay vertices, $\Dt \simeq {(z_{\rm Rec} - z_{\rm Tag})}/{\beta \gamma c} \equiv {\Delta z}/{\beta \gamma c}$. To obtain the \Dt\ distribution, we reconstruct the tag-side vertex from the tracks not used to reconstruct \Brec~\cite{ResFunc} and employ the flavour tagging routine described in Ref.~\cite{Tagging}. The tagging information is represented by two parameters, the \Btag\ flavour, $q$ and $r$. The parameter, $r$, is an event-by-event, MC determined flavour-tagging dilution factor that ranges from $r = 0$ for no flavour discrimination to $r = 1$ for unambiguous flavour assignment.

\section{TIME-DEPENDENT $CP$ VIOLATION MEASUREMENT IN \Kzpiz}
Direct $CP$ violation has been observed in $\Bz \rightarrow \Kp \pim$~\cite{kpi_exp} and is found to be significantly different from that in $B^{\pm} \rightarrow K^{\pm} \pi^{0}$. This unexpected result may indicate the presence of new physics (NP) or poor understanding of strong interaction effects in $B$ decays. A model-independent test for NP is possible via an isospin sum rule with high statistics~\cite{kpi} which gives a relation among the direct $CP$ violation asymmetries measured in all possible $B \to K \pi$ modes, i.e. $B^0 \to K^0 \pi^0$, $K^+ \pi^-$, $B^+ \to K^+ \pi^0$ and $K^0 \pi^+$,
\begin{equation}
\mathcal{A}_{K^+\pi^-} + \mathcal{A}_{K^0\pi^+}\frac{\mathcal{B}(K^0\pi^+)\tau_{B^0}}{\mathcal{B}(K^+\pi^-)\tau_{B^+}} = \mathcal{A}_{K^+\pi^0}\frac{2\mathcal{B}(K^+\pi^0)\tau_{B^0}}{\mathcal{B}(K^+\pi^-)\tau_{B^+}} +\mathcal{A}_{K^0\pi^0}\frac{2\mathcal{B}(K^0\pi^0)}{\mathcal{B}(K^+\pi^-)}.
\end{equation}
Here, $\mathcal{B}$ represents the branching ratio of each decay mode and $\tau_{B^+}$ is the lifetime of the charged $B$ meson. The sum rule's theoretical precision is determined by SU(2) flavour symmetry, i.e. a few \%, therefore the sum rule provides a clean test for new physics. As a fundamental input, the $CP$ violation measurement in $B^0 \to K^0 \pi^0$ is currently the least known, experimentally. Since the branching fractions and $CP$ asymmetries of other $B \to K \pi$ decay modes have been measured to good precision~\cite{PDG}, $\mathcal{A}_{K^0\pi^0}$ is predicted with a small error. In addition to the $B^0 \to K^0_S \pi^0$ mode, we measure $CP$ asymmetry in $B^0 \to K^0_L \pi^0$ decay for the first time, in order to maximize sensitivity to the direct $CP$ violation parameter, $\mathcal{A}_{K^0\pi^0}$.

The signal yield of \Kspiz\ events is found from a three-dimensional extended unbinned maximum likelihood fit to \Mbc, \De\ and \LR\ to be $657\pm37$ events. For \Klpiz, \Mbc\ can be calculated from the direction of the \Kl\ cluster while \De\ cannot be calculated. The signal yield of $285\pm52$ events is extracted from a two-dimensional fit to \Mbc\ and \LR\ and has a significance of 3.7$\sigma$ including systematic uncertainties.

\begin{figure*}[t]
\centering
\includegraphics[width=135mm]{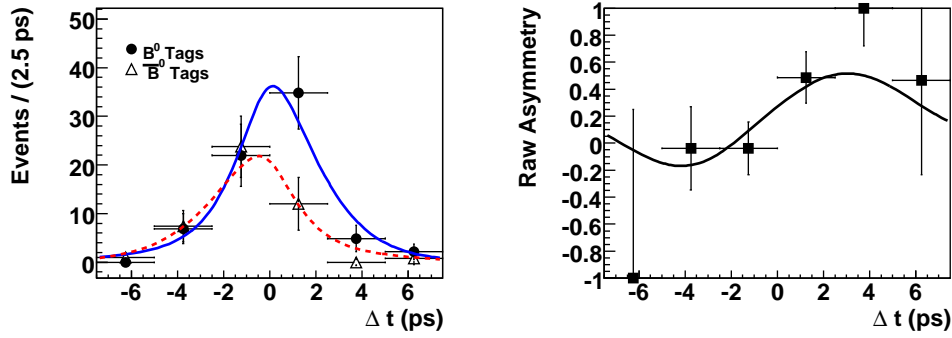}
\caption{The left plot shows the background subtracted fit to \Dt\ for the good tags region, $0.5 < r \leq 1.0$, where the solid (dashed) curve represents \Bz\ (\Bzb) tags. The right plot shows the \BzBzb\ raw asymmetry, $(N_{\Bz} - N_{\Bzb})/(N_{\Bz} + N_{\Bzb})$, where $N_{\Bz} \, (N_{\Bzb})$ is the number of signal \Bz\ (\Bzb) tags in \Dt.} \label{fig_kspi0}
\end{figure*}

As the vertex position of \Klpiz\ cannot be determined and the vertex reconstruction efficiency of \Kspiz\ is $\sim33\%$ due to the long lifetime of the \Ks, these events can still be used to calculate \Acp\ by integrating Eq.~\ref{eq_tcpv} over \Dt. We extract the $CP$ parameters,
\begin{eqnarray}
\Acp &=& +0.14 \pm 0.13 \; \mbox{(stat)} \pm 0.06 \; \mbox{(syst)},\nonumber \\
\Scp &=& +0.67 \pm 0.31 \; \mbox{(stat)} \pm 0.08 \; \mbox{(syst)}
\end{eqnarray}
and the fit results for the \Dt\ component is shown in Fig.~\ref{fig_kspi0}. We find that the mixing induced component is consistent with charmonium, $\Scp(b \rightarrow c \bar c s) = 0.681 \pm 0.025$~\cite{HFAG}, and that there is a $1.9\sigma$ deviation between our measurement of \Acp\ and the expectation from the isospin sum rule, $\Acp = -0.17\pm0.06$~\cite{kpi}.

\section{TIME-DEPENDENT DALITZ PLOT $CP$ VIOLATION MEASUREMENT IN \Kspipi}
The signal yield of \Kspipi\ events is found from a one-dimensional fit to \De\ to be $1944 \pm 98$ events and the resonances considered in the signal model (Eq. \ref{eq_sigmod}) are the \Kspm, \Kstarpm, \rhoz, \fz, \ftwo, \fX\ and a non-resonant component. We obtain the $CP$ parameters,

\vspace{5pt}
\begin{minipage}[]{0.5\columnwidth}
  \begin{tabbing}
    Solution 1: \= $-2 \log {\cal L} = 18472.5$\\
    \> $f(\Kstarp \pim) = 61.7 \pm 10.4\%$
  \end{tabbing}
  \vspace{-15pt}
  \begin{eqnarray}
    \Acp(\rhoz \Ks) &=& +0.03^{+0.23}_{-0.24} \pm 0.11 \pm 0.10, \nonumber \\
    \phione(\rhoz \Ks) &=& (+20.0^{+8.6}_{-8.5} \pm 3.2 \pm 3.5)^{\circ}, \nonumber \\
    \Acp(\fz \Ks) &=& -0.06 \pm 0.17 \pm 0.07 \pm 0.09, \nonumber \\
    \phione(\fz \Ks) &=& (+12.7^{+6.9}_{-6.5} \pm 2.8 \pm 3.3)^{\circ}, \nonumber
  \end{eqnarray}
\end{minipage}
\begin{minipage}[]{0.47\columnwidth}
  \begin{tabbing}
    Solution 2: \= $-2 \log {\cal L} = 18465.0$\\
    \> $f(\Kstarp \pim) = 17.4 \pm 5.0\%$
  \end{tabbing}
  \vspace{-15pt}
  \begin{eqnarray}
    \Acp(\rhoz \Ks) &=& -0.16 \pm 0.24 \pm 0.12 \pm 0.10, \nonumber \\
    \phione(\rhoz \Ks) &=& (+22.8 \pm 7.5 \pm 3.3 \pm 3.5)^{\circ}, \nonumber \\
    \Acp(\fz \Ks) &=& +0.00 \pm 0.17 \pm 0.06 \pm 0.09, \nonumber \\
    \phione(\fz \Ks) &=& (+14.8^{+7.3}_{-6.7} \pm 2.7 \pm 3.3)^{\circ},
  \end{eqnarray}
\end{minipage}

\begin{figure*}[t]
  \centering
  \includegraphics[width=65mm]{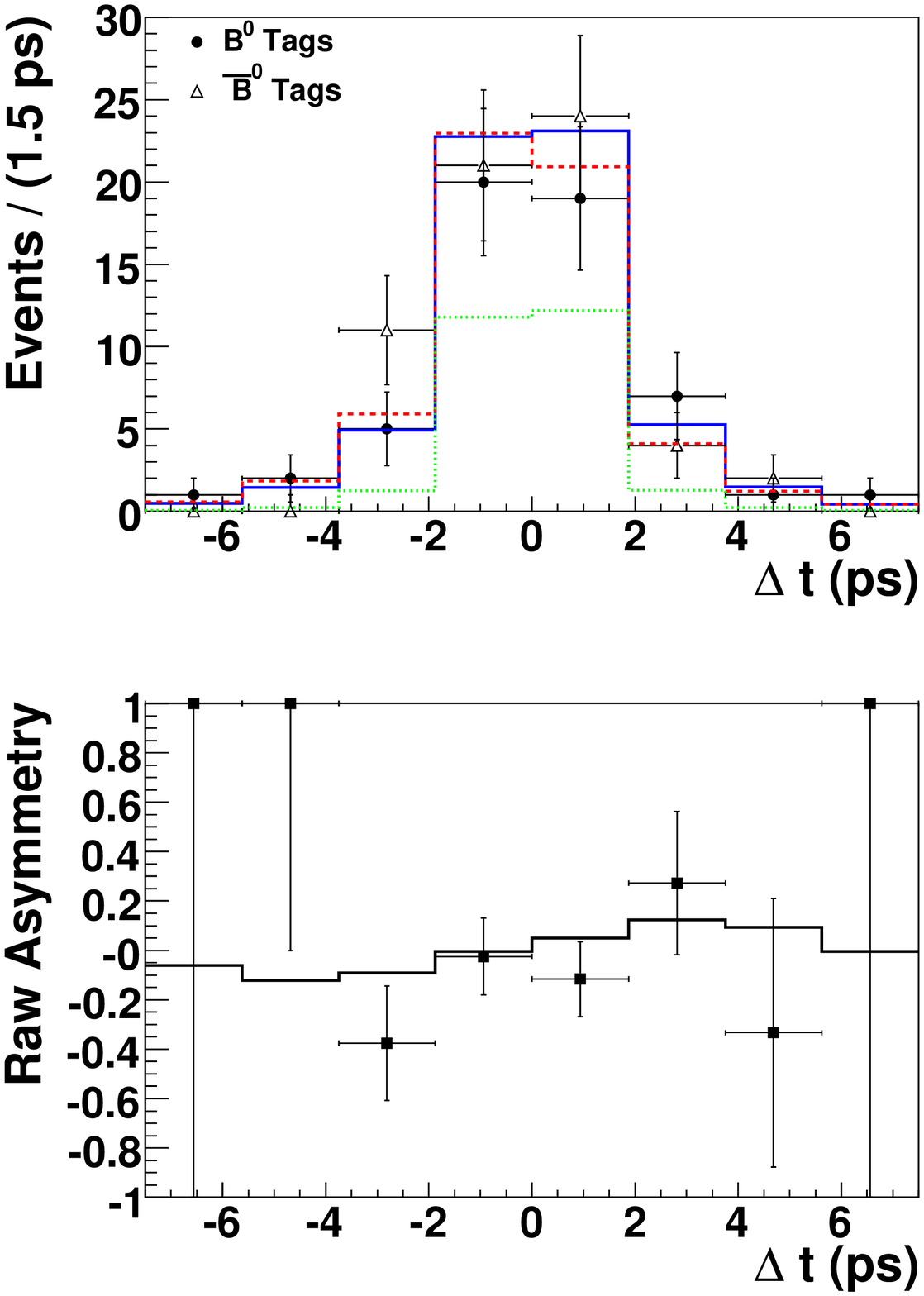}
  \includegraphics[width=65mm]{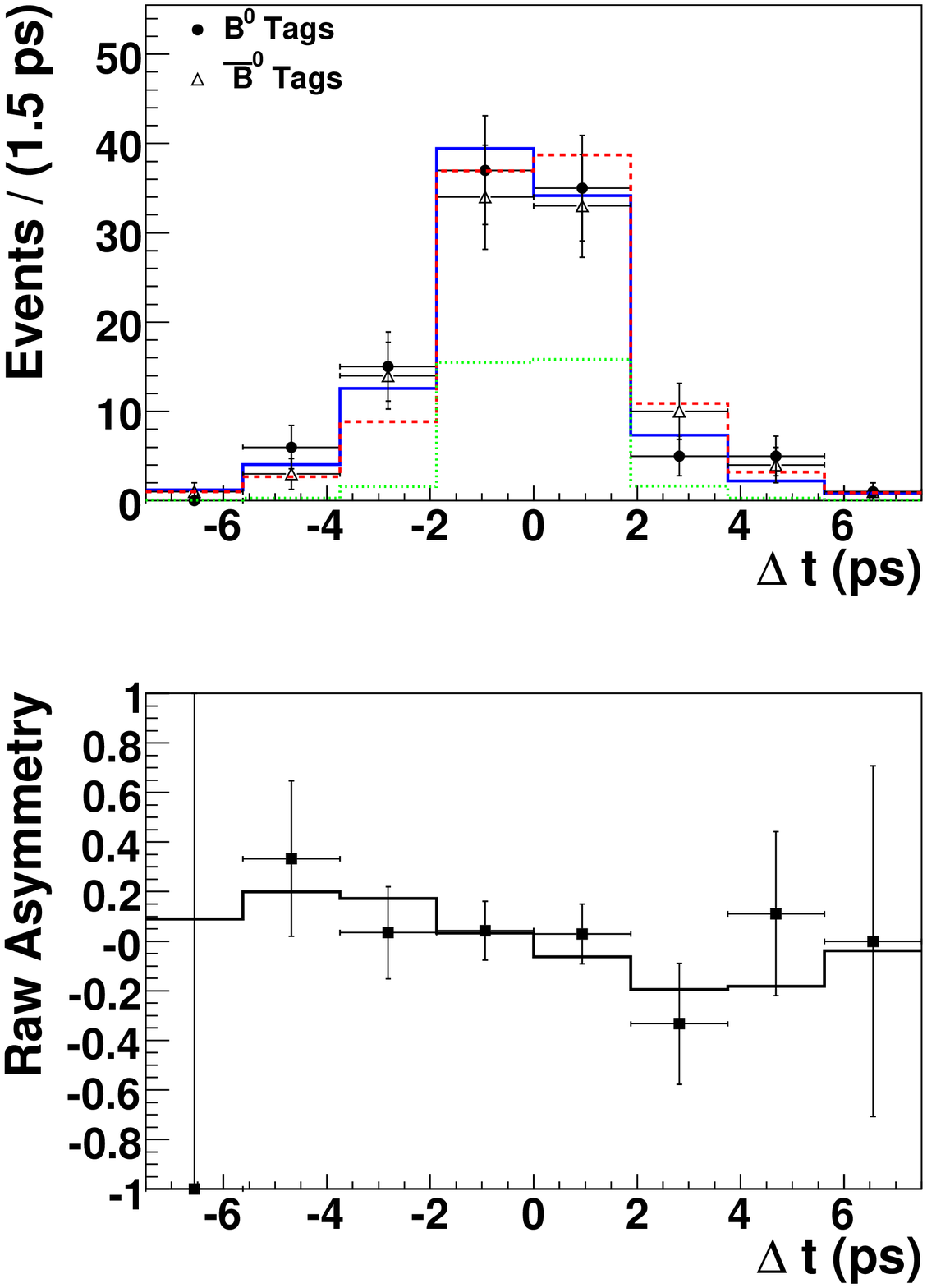}
  \put(-223,216){(a)}
  \put(-36,216){(b)}
  \caption{Time-dependent Dalitz plot fit results for \Kspipi\ in (a), the \rhoz\ region and (b), the \fz\ region. The top plots show the \Dt\ distribution for \Bz\ (solid) and \Bzb\ (dashed) tags. These plots contain only good tags, $0.5 < r \leq 1.0$ and the dotted curve represents the background contribution. The bottom plots show the \BzBzb\ raw asymmetry, $(N_{\Bz} - N_{\Bzb})/(N_{\Bz} + N_{\Bzb})$, where $N_{\Bz} \, (N_{\Bzb})$ is the number of \Bz\ (\Bzb) tags in \Dt.}
  \label{fig_kspipi}
\end{figure*}

\vspace{5pt}
where the first error is statistical, the second is systematic and the third is the Dalitz plot signal model uncertainty. Figure~\ref{fig_kspipi} shows the fit results. The high $\Kstarp \pim$ fraction of Solution 1 is in agreement with some phenomenological estimates~\cite{Kstarpi} and may also be favoured by the total $K-\pi$ $S$-wave phase shift when compared with that measured by LASS~\cite{LASS}.  As the likelihood difference is not found to be significant, we do not rule between these solutions where \Acp\ is consistent with null asymmetry and \phione\ agrees with charmonium.

\section{TIME-DEPENDENT DALITZ PLOT $CP$ VIOLATION MEASUREMENT IN \Kskk}
\begin{figure*}[t]
  \centering
  \includegraphics[width=65mm]{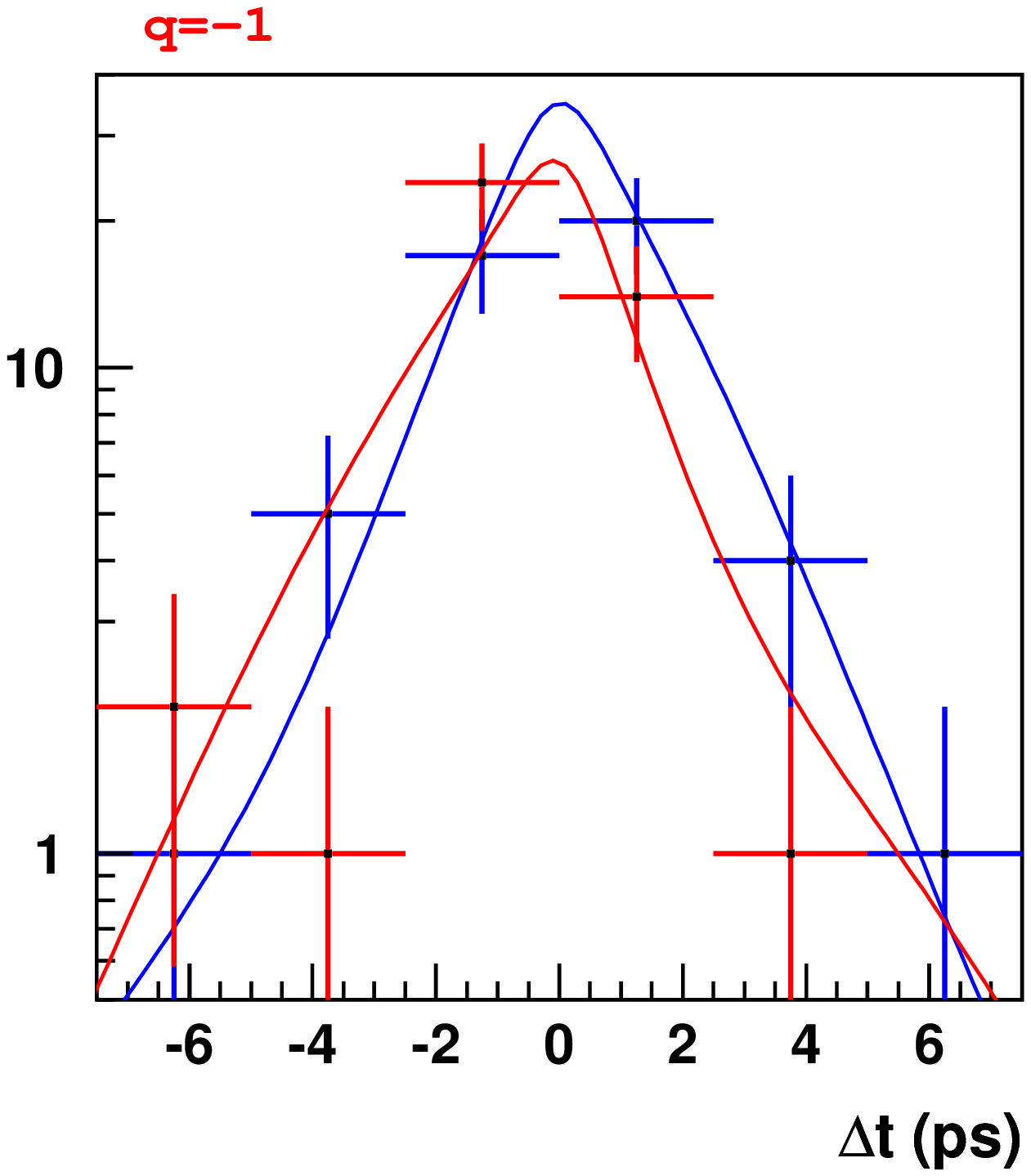}
  \includegraphics[width=65mm]{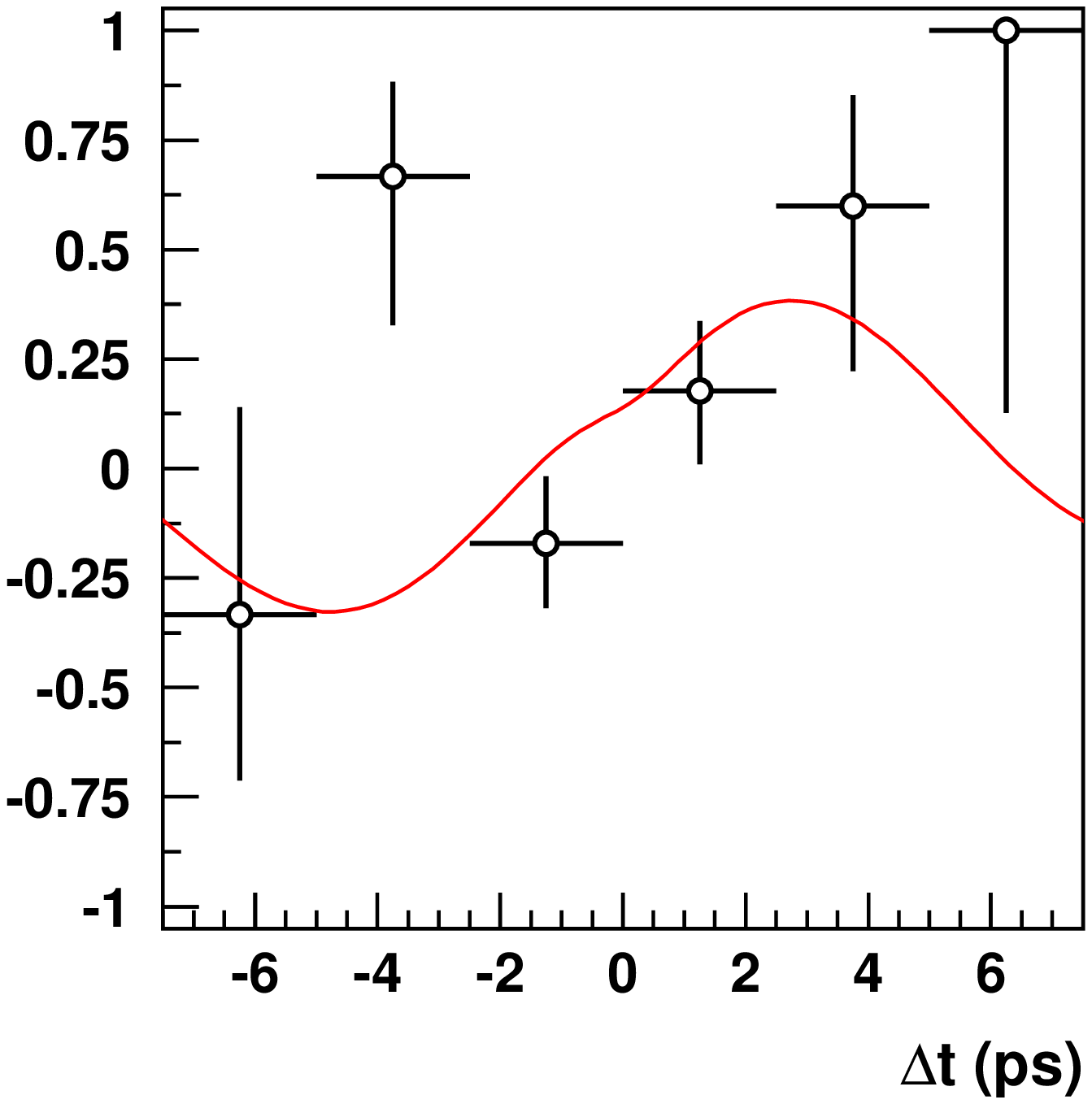}
  \caption{Time-dependent Dalitz plot fit results for \Kskk\ in the $\phi(1020)$ region. The left plot show the \Dt\ distribution for \Bz\ (blue) and \Bzb\ (red) tags for good tags, $0.5 < r \leq 1.0$. The right plot shows the \BzBzb\ raw asymmetry}
  \label{fig_kskk}
\end{figure*}

The signal yield of \Kskk\ is extracted to be $1269 \pm 51$ events from a two-dimensional fit to \Mbc\ and \De\ in each $r$-bin. We consider the \fz, $\phi(1020)$, $f_{X}(1500)$, \chic, and a non-resonant component in the signal model. Four solutions with similar likelihood are found where two solutions arise from the interference between \fz\ and the non-resonant component and the other two from $f_{X}(1500)$ and the non-resonant component. Using external information from \Kspipi, if the $f_{X}(1500)$ is the $f_{0}(1500)$ for both \Kspipi\ and \Kskk, the ratio of branching fractions, ${\cal B}(f_{0}(1500) \rightarrow \pip \pim) / {\cal B}(f_{0}(1500) \rightarrow \Kp \Km)$, prefers the solution with the low $f_{X}(1500) \Ks$ fraction. Similarly the ratio, ${\cal B}(\fz \rightarrow \pip \pim) / ({\cal B}(\fz \rightarrow \pip \pim) + {\cal B}(\fz \rightarrow \Kp \Km))$ prefers the solution with the low $\fz \Ks$ fraction. The preferred solution is,
\begin{eqnarray}
  \Acp(\fz \Ks) \!\!&=&\!\! -0.02 \pm 0.34 \pm 0.08 \pm 0.09 \nonumber \\
  \phione(\fz \Ks) \!\!&=&\!\! (28.2^{+9.8}_{-9.9} \pm 2.0 \pm 2.0)^{\circ} \nonumber \\
  \Acp(\phi(1020) \Ks) \!\!&=&\!\! +0.31^{+0.21}_{-0.23} \pm 0.04 \pm 0.09 \nonumber \\
  \phione(\phi(1020) \Ks) \!\!&=&\!\! (21.2^{+9.8}_{-10.4} \pm 2.0 \pm 2.0)^{\circ}
\end{eqnarray}
where the first error is statistical, the second is systematic and the third is the Dalitz plot signal model uncertainty. Figure~\ref{fig_kskk} shows the fit results where \Acp\ is consistent with null asymmetry and \phione\ agrees with charmonium.

\begin{acknowledgments}
We thank the KEKB group for excellent operation of the
accelerator, the KEK cryogenics group for efficient solenoid
operations, and the KEK computer group and
the NII for valuable computing and SINET3 network
support.  We acknowledge support from MEXT and JSPS (Japan);
ARC and DEST (Australia); NSFC (China); 
DST (India); MOEHRD and KOSEF (Korea); 
KBN (Poland); MES and RFAAE (Russia); ARRS (Slovenia); SNSF (Switzerland); 
NSC and MOE (Taiwan); and DOE (USA).
\end{acknowledgments}



\begin{thebibliography}{99} 

\bibitem{C}
  N.~Cabibbo, Phys. Rev. Lett. {\bf 8}, 214 (1964).
\bibitem{KM}
  M.~Kobayashi and T.~Maskawa, Prog. Theor. Phys. {\bf 49}, 652 (1973).
\bibitem{HFAG}
  E. Barberio {\it et al.} (Heavy Flavor Averaging Group), 
  arXiv:0808.1297 [hep-ex] and online update for Winter 2008 at 
  http://www.slac.stanford.edu/xorg/hfag.
\bibitem{Sanda}
  A.~B.~Carter and A.~I.~Sanda, Phys. Rev. Lett. {\bf 45}, 952 (1980); 
  A.~B.~Carter and A.~I.~Sanda, Phys. Rev. D {\bf 23}, 1567 (1981); 
  I.~I.~Bigi and A.~I.~Sanda, Nucl. Phys. {\bf 193}, 85 (1981).
\bibitem{Belle}
  A.~Abashian {\it et al.} (Belle Collab.),
  Nucl. Instr. and Meth. A {\bf 479}, 117 (2002).
\bibitem{KEKB}
  S.~Kurokawa and E.~Kikutani, Nucl. Instr. and Meth. A {\bf 499}, 1 (2003),
  and other papers included in this volume.
\bibitem{SFW}
  K. Abe {\it et al.} (Belle Collaboration), Phys. Rev. Lett. {\bf 87},  101801 (2001);
  K. Abe {\it et al.} (Belle Collaboration), Phys. Lett. {\bf B 511}, 151 (2001);
  S. H. Lee, K. Suzuki, {\it et al.} (Belle Collaboration), Phys. Rev. Lett. {\bf 91}, 261801 (2003).
\bibitem{ResFunc}
  H.~Tajima {\it et al.} Nucl.~Instr.~and~Meth.~A {\bf 533}, 370 (2004).
\bibitem{Tagging}
  H.~Kakuno {\it et al.}, Nucl. Instr. and Meth. A {\bf 533}, 516 (2004).
\bibitem{kpi_exp}
  B.~Aubert {\it et al.} (BaBar Collab.), Phys. Rev. Lett. {\bf 99}, 021603 (2007);
  S.-W.~Lin {\it et al.} (Belle Collab.), Nature {\bf 452}, 332 (2008).
\bibitem{kpi}
  M.~Gronau, Phys. Lett. B {\bf 627}, 82 (2005).
\bibitem{PDG}
  C.~Amsler (Particle Data Group), Phys. Lett. B {\bf 567}, 1 (2008).
\bibitem{Kstarpi}
  V.~L.~Chernyak, Phys. Lett. B {\bf 509}, 273 (2001).
\bibitem{LASS}
  D.~Aston {\it et al.} (LASS Collab.), Nucl. Phys. B {\bf 296}, 493 (1988).

\end{thebibliography}
\end{document}